\documentclass[prl,aps,twocolumn,showpacs,preprintnumbers,amsmath,amssymb]{revtex4}
\usepackage{dcolumn}
\usepackage{graphicx}

\begin{document}

\title{
Mott physics in $2p$ electron dioxygenyl magnet : 
O$_{2}$$M$F$_{6}$ ($M$=Sb, Pt)
}
\author{Minjae Kim and B. I. Min}
\affiliation{Department of Physics, PCTP,
	Pohang University of Science and Technology, Pohang, 790-784, Korea}
\date{\today}

\begin{abstract}
We have investigated electronic structures and magnetic properties
of O$_{2}$$M$F$_{6}$ ($M$=Sb, Pt), which are composed of
two building blocks of strongly correlated electrons:
O$_{2}^{+}$ dioxygenyls and $M$F$_{6}^{-}$ octahedra, by
employing the first-principles electronic structure band method.
For O$_{2}$SbF$_{6}$, as a reference system of O$_{2}$PtF$_{6}$,
we have shown that the Coulomb correlation of O(2$p$) electrons
drives the Mott insulating state.
For O$_{2}$PtF$_{6}$, we have demonstrated that the Mott insulating
state is induced by the combined effects of the
Coulomb correlation of O(2$p$) and Pt(5$d$) electrons
and the spin-orbit (SO) interaction of Pt(5$d$) states.
The role of the SO interaction in forming the Mott insulating state of
O$_{2}$PtF$_{6}$ is similar to the case of Sr$_{2}$IrO$_{4}$
that is a prototype of a SO induced Mott system with J$_{eff}=1/2$.
\end{abstract}


\maketitle


Mott transition in strongly correlated electron systems has
been extensively studied for 3$d$ transition metal (TM) oxides,
in which the degenerate 3$d$ states are usually split
by the crystal-field (CF) and then the Coulomb correlation effect
produces the energy separation
between the upper and lower Hubbard bands.\cite{imada}
The spin-orbital dependent superexchange interaction of Kugel-Khomskii
type can be derived based on orbital polarizations of 3$d$ states
in these Mott insulating states.\cite{Kugel,Bhkim10}
According to recent experimental and theoretical works,
the similar Mott physics is realized
in 2$p$ electron molecular solids (KO$_{2}$)\cite{solovyev,mkim,kang}
and also
in 5$d$ TM oxides (Sr$_{2}$IrO$_{4}$)\cite{bkim,moon,pesin,watanabe},
which signifies a new paradigm of spin-orbital physics.
The bands in 2$p$ electron molecular solids are almost
molecular-level like  due to the weak intermolecular interaction,
and so the band width $W$ near the Fermi level $E_F$ is small
with respect to on-site Coulomb repulsion $U$,
resulting in the Mott insulating state.\cite{solovyev,mkim,kang}
On the other hand, 
in 5$d$ TM oxides, the large spin-orbit (SO) interaction as much as
$\sim$0.4 eV\cite{bkim} brings about the small $W$
by lifting the degeneracy of 5$d$ states and also by reducing
the hopping strength.\cite{bkim,watanabe}
In the strong SO interaction limit,
electronic structures of 5$d$ TM oxides are determined
by the relative magnitude between $W$
and $U$ of 5$d$ electrons.\cite{moon,pesin}
In this context, dioxygenyl magnet O$_{2}$PtF$_{6}$ is interesting
in that it is a 2$p$ electron molecular solid containing 5$d$ TM element,
which possesses both effects of the strong Coulomb correlation
and the strong SO interaction.

\begin{figure}[b]
\includegraphics[width=8cm]{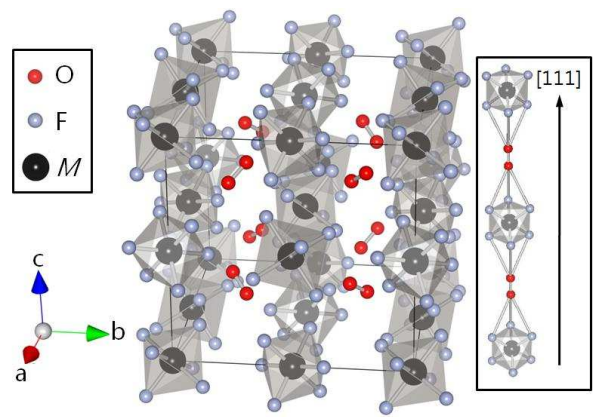}
\caption{(Color online) Crystal structure of O$_{2}$$M$F$_{6}$ ($M$=Sb, Pt)
composed of O$_{2}^{+}$ dioxygenyls and $M$F$_{6}^{-}$ octahedra.
There are eight O$_{2}^{+}$ and $M$F$_{6}^{-}$ ions in a unit cell.
In the right, the molecular configuration of O$_{2}^{+}$ and
$M$F$_{6}^{-}$ along the [111] direction is shown.
}
\label{fig1}
\end{figure}

\begin{figure}[t]
\includegraphics[width=8.5cm]{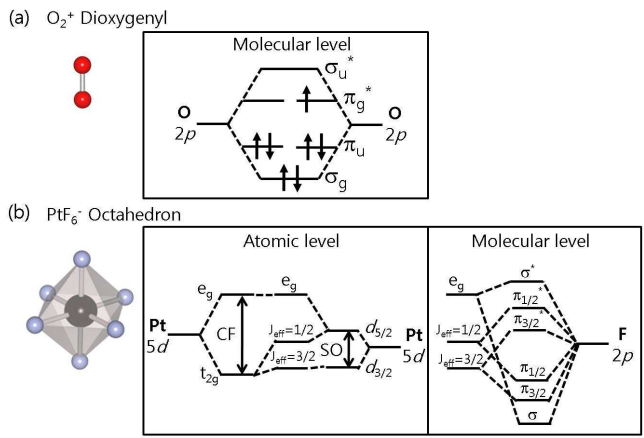}
\caption{(Color online) (a) Molecular level of O$_{2}^{+}$
dioxygenyl ion. (b) Schematic local electronic structure of PtF$_{6}^{-}$
ion. Pt(5$d$) atomic states are split into e$_{g}$ and
J$_{eff}=1/2,3/2$ under the octahedral crystal field (CF)
with the spin-orbit (SO) interaction. Then Pt(5$d$)-F(2$p$)
hybridization splits those levels into bonding and antibonding
molecular states with $\pi$ and $\sigma$ channels.
}
\label{fig2}
\end{figure}

O$_{2}$$M$F$_{6}$ ($M$=Sb, Pt) were reported to be
insulators.\cite{grill,disalvo,botkovitz}
O$_{2}$$M$F$_{6}$ is composed of O$_{2}^{+}$ dioxygenyls
and $M$F$_{6}^{-}$ octahedra, as depicted in Fig.~\ref{fig1}.
Dioxygenyl O$_{2}^{+}$ ion has the electronic configuration of
$\sigma_{g}^{2}\pi_{u}^{4}\pi_{g}^{*1}$ (see Fig.~\ref{fig2}(a)).
Hence the localized magnetic moment would be generated
from one unpaired electron on the degenerate $\pi_{g}^{*}$ orbital.
Magnetic susceptibility data showed
the Curie-Weiss behaviors with low Curie temperatures,
$T_C =0.8$ K for O$_{2}$SbF$_{6}$ and
$T_C =4$ K for O$_{2}$PtF$_{6}$, respectively,
which implies the rather weak superexchange interaction
between localized magnetic moments.
Electron spin resonance (ESR) experiment also revealed
the existence of the localized magnetic moment of O$_{2}^{+}$ in O$_{2}$SbF$_{6}$
and the ferrimagnetic ordering between the magnetic moments of O$_{2}^{+}$ and
Pt in O$_{2}$PtF$_{6}$.\cite{disalvo}

Local electronic structure of O$_{2}^{+}$ in Fig.~\ref{fig2} is
similar to that of O$_{2}^{-}$ in KO$_{2}$ superoxide
that has one hole (three electrons) in the $\pi_{g}^{*}$ state.
Electronic structure calculations for various superoxides
(KO$_{2}$, RbO$_{2}$, Rb$_{4}$O$_{6}$) indicated
that the conventional local density approximation (LDA)
or the generalized gradient approximation (GGA)
underestimates the Coulomb correlation interaction
of open shell 2$p$ electrons in superoxides.
Only after the inclusion of on-site Coulomb repulsion $U$ in the LDA+$U$
or GGA+$U$ schemes, the insulating nature of superoxides was
properly described.\cite{solovyev,mkim,Attema05,kovacik,winterlik,Ylvi,nandy}

We have investigated electronic structures and
magnetic properties of O$_{2}$PtF$_{6}$ by employing the first-principles
band structure calculation incorporating both the Coulomb correlation
and the SO effect. As a reference system, we have also examined
electronic structures of O$_{2}$SbF$_{6}$.
For O$_{2}$SbF$_{6}$, we have confirmed that the Coulomb correlation of
O(2$p$) electrons yields the Mott insulating state,
consistently with experiments.
For O$_{2}$PtF$_{6}$, we have shown that not only correlation effects of
both O(2$p$) and Pt(5$d$) electrons but also the strong SO interaction
of the 5$d$ states are essential to describe its Mott insulating state.

We have performed electronic structure calculations
for the experimental unit cell of O$_{2}$$M$F$_{6}$
($M$=Sb, Pt) with bcc structure of
$Ia_3$(206) space group.\cite{graudejus}
We have employed the full-potential augmented plane
wave (FLAPW) band method\cite{FLAPW}
implemented in WIEN2k package.\cite{Blaha}
We have considered the on-site $U$ of both O($2p$) and Pt(5$d$) electrons
in the GGA+$U$\cite{U}
and the SO interaction effect is included as a second variational procedure
in the GGA+SO and the GGA+SO+$U$ scheme.
The spin direction has chosen to be (001) direction
for the consideration of SO interaction.
For results below, $U$ value was chosen to be 10 eV
for both O($2p$)  and Pt(5$d$) states.\cite{Uval,takeshige}
We have assumed the ferromagnetic state for O$_{2}$SbF$_{6}$,
and the ferrimagnetic ordering between the magnetic moments of
O$_{2}^{+}$ and Pt for O$_{2}$PtF$_{6}$.\cite{mconf}
To allow the orbital polarization
of $\pi_{g}^{*}$ states, as was realized in KO$_{2}$ superoxide,\cite{mkim}
we have removed all the symmetry operators except inversion symmetry.


\begin{figure}[t]
\includegraphics[width=8.0cm]{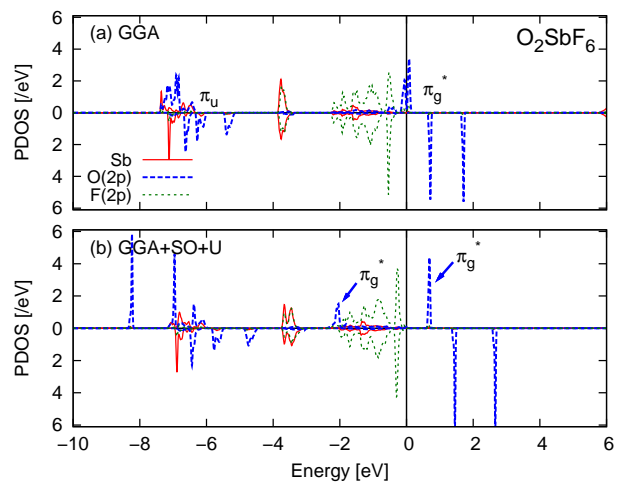}
\caption{(Color online)
(a) Partial density of states (PDOS) of O$_{2}$SbF$_{6}$ in the GGA.
The red solid, blue dashed, and green dotted lines denote
DOS of Sb, O($2p$), and F(2$p$) states, respectively.
(b) PDOS in the GGA+SO+$U$. The inclusion of Coulomb correlation yields
the separation of $\pi_{g}^{*}$ band into the occupied lower Hubbard
and unoccupied upper Hubbard bands, as indicated by blue arrows.
$U$ value was chosen to be 10 eV for O($2p$).
}
\label{fig3}
\end{figure}

Figure~\ref{fig3} shows the calculated density of states (DOS) of
O$_{2}$SbF$_{6}$ both in the GGA and GGA+SO+$U$ schemes.
Most valence electrons of Sb are transferred into F, and
4$d^{10}$ occupied states of Sb$^{5+}$ are located $\sim$25 eV below $E_F$.
Hence only a small amount of Sb states originating from the hybridization
with F(2$p$) states exists near $E_F$.
Therefore, O$_{2}$SbF$_{6}$ is a typical O($2p$) molecular solid
like KO$_2$.
The GGA in Fig.~\ref{fig3}(a) yields the metallic state
for O$_{2}$SbF$_{6}$ with $\pi_{g}^{*}$ spin $\uparrow$ states
being located at $E_F$.
On the other hand, the GGA+SO+$U$ in Fig.~\ref{fig3}(b)
considering the correlation effect
of O($2p$) electrons yields the correct insulating state by separating the
$\pi_{g}^{*}$ states into occupied (lower Hubbard)  and unoccupied
(upper Hubbard) states with a gap of $\Delta_H=2.55$ eV.
We have also found that electronic structures
in the GGA+$U$ are almost the same as those in the GGA+SO+$U$,
implying that the role of the SO interaction in O$_{2}$SbF$_{6}$ is minor.
Note that, in KO$_{2}$ superoxide, the degenerate $\pi_{g}^{*}$
states tend to be split by coherent tilting of the O$_{2}^{-}$ molecular axes.
This mechanism is called  "magnetogyration", which invokes
the structural and magnetic transitions concomitantly.\cite{kanzig}
Likewise, the degeneracy of the $\pi_{g}^{*}$ states in O$_{2}$SbF$_{6}$
is lifted by trigonal CF coming from neighboring F$^{-}$
ions.\cite{grill,disalvo}.
This feature is revealed in the spin-density (SD) plot
in Fig.~\ref{fig5}(a), which manifests
the directionally polarized $\pi_{g}^{*}$ orbital shape
at each O$_{2}^{+}$  site.

There was controversy on the existence of SO effects
in the $\pi_{g}^{*}$ states of O$_{2}$$M$F$_{6}$.
In one group,\cite{grill,shamir}
$g_{eff}$ factor determined from the magnetic susceptibility and ESR
was close to $2$ independently of the direction of magnetic field,
which reflects the negligible SO effect,
while, in another,\cite{disalvo}
$g_{eff}$ was reduced a lot along the molecular axis (1.73),
which reflects the large orbital moment along that molecular axis.
In the present GGA+SO+$U$ calculation, the orbital magnetic moment
is indeed induced along the molecular axis of O$_{2}^{+}$.
However, its magnitude, 0.02 $\mu_{B}$
per each O$_{2}^{+}$, is only $\sim2\%$ of the
spin magnetic moment (1.00 $\mu_{B}$), which is too small to produce
reduced $g$ factor as reported by Ref.\cite{disalvo}.
This result implies that the SO interaction is suppressed
in O$_{2}$SbF$_{6}$ due to the quenched orbital moment of O$_{2}^{+}$,
similarly to the case of low symmetry structure of 
KO$_{2}$.\cite{mkim,nandy}

\begin{figure}[t]
           \includegraphics[width=8.0cm]{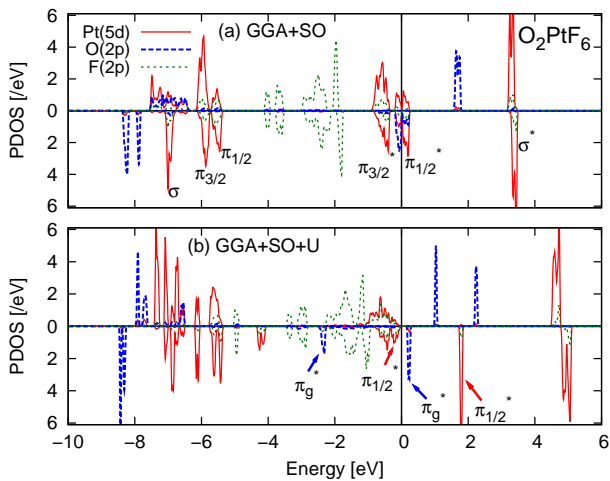}
\caption{(Color online)
(a) PDOS of O$_{2}$PtF$_{6}$ in the GGA+SO.
The red solid, blue dashed, and green dotted lines denote
DOS of Pt(5$d$), O($2p$), and F(2$p$) states, respectively.
(b) In GGA+SO+$U$,
inclusion of the Coulomb correlation and the SO interaction
yields the separation of
(i) $\pi_{g}^{*}$ states of O($2p$), as indicated by blue arrows,
and (ii) $\pi_{1/2}^{*}$ states of Pt($5d$), as indicated by red arrows.
$U$ values were chosen to be 10 eV for both O($2p$) and Pt($5d$).
}
\label{fig4}
\end{figure}

Figure~\ref{fig4} shows the calculated DOS of O$_{2}$PtF$_{6}$.
Noteworthy is that Mott insulating state
is obtained only in the GGA+SO+$U$ (Fig.~\ref{fig4}(b)),
in which both the Coulomb correlations of O($2p$) and Pt($5d$)
electrons and the SO interaction are considered.
Namely, the GGA, GGA+SO, and GGA+$U$ schemes
without including the SO interaction
do not give correct insulating state of O$_{2}$PtF$_{6}$.
Local electronic structure of Pt$^{5+}$ in O$_{2}$PtF$_{6}$ is
close to that of Ir$^{4+}$ in Sr$_{2}$IrO$_{4}$.
Both Pt$^{5+}$ and Ir$^{4+}$ have 5$d^{5}$ electronic configuration
under the octahedral CF with strong SO interaction (see Fig.~\ref{fig2}(b)).
Since the CF is stronger than the Hund exchange interaction,
they both have the low-spin (LS) states.
As shown in Fig.\ref{fig2}(b), the SO interaction splits $t_{2g}$
states of Pt into J$_{eff}=1/2,3/2$, which then hybridize with F(2$p$) states
to produce $\pi_{1/2}^{*}$ states at $E_F$ (see Fig.~\ref{fig4}(a)).
Due to the shorter Pt-F bond length ($\sim$1.88 ${\AA}$)\cite{graudejus}
than that of Ir-O ($\sim$2.00 ${\AA}$),\cite{shimura}
the bands in O$_{2}$PtF$_{6}$ are almost molecular-level like.
In the GGA+SO in Fig.~\ref{fig4}(a),
the SO split $\pi_{1/2}^{*}$ states are still degenerate
to make O$_{2}$PtF$_{6}$ metallic.
The degenerate $\pi_{1/2}^{*}$ states become split
only after the inclusion of Coulomb interaction of Pt(5$d$) electrons,
as shown in Fig.\ref{fig4}(b).
Note that the $\pi_{1/2}^{*}$ states are not to be
split in the GGA+$U$ even with very large $U$ value,
unless the SO interaction is included.

\begin{figure}[t]
           \includegraphics[width=7cm]{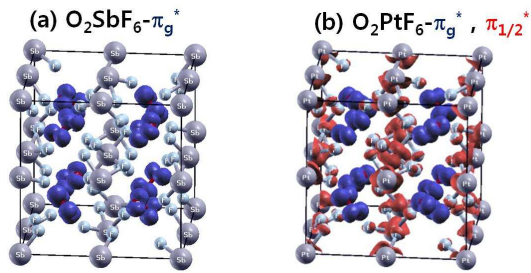}\\
\caption{(Color online)
(a) SD plot of O$_{2}$SbF$_{6}$ in the GGA+SO+$U$.
Directionally polarized $\pi_{g}^{*}$ orbitals in blue are manifested.
(b) SD plot of O$_{2}$PtF$_{6}$ in the GGA+SO+$U$.
The directionally polarized $\pi_{g}^{*}$ orbitals in blue
and the directionally non-polarized  $\pi_{1/2}^{*}$ orbitals in red
are manifested.
}
\label{fig5}
\end{figure}

In general, ions with $t_{2g}^{5}$ occupation in the LS state are
Jahn-Teller (JT) active. It was once reported that reduced $g_{eff}$ factor
($g=1.86$) for the LS state ($S=1/2$) of Pt$^{5+}$ in O$_{2}$PtF$_{6}$
could be explained by the JT effect.\cite{disalvo}
The JT distortion, however, has not been detected in O$_{2}$PtF$_{6}$.
The neutron diffraction\cite{ibers} provided that PtF$_{6}^{-}$
has a regular octahedron structure.
No anomaly in the temperature dependent magnetic susceptibility\cite{disalvo}
for O$_{2}$PtF$_{6}$ also indicates the absence of
structural phase transition.
Electronic structures in Fig.~\ref{fig4} really verify that
the degeneracy of $t_{2g}$ states is lifted by the SO effect
not by the JT effect.
The orbital magnetic moment of Pt ion is as much as 0.52 $\mu_{B}$,
which is comparable to the spin magnetic moment of 0.60 $\mu_{B}$.
Thus, it is this large orbital magnetic moment of Pt that is responsible
for the deviation of the observed $g_{eff}$ factor from two.
Indeed, the feature of the lifted degeneracy is revealed in the SD plot
in Fig.~\ref{fig5}(b), which manifests
the directionally non-polarized $\pi_{1/2}^{*}$ orbital shape
at each Pt$^{5+}$ site,
in contrast to the directionally polarized $\pi_{g}^{*}$ orbital shape
at each O$_{2}^{+}$  site.

\begin{figure}[t]
           \includegraphics[width=7.2cm]{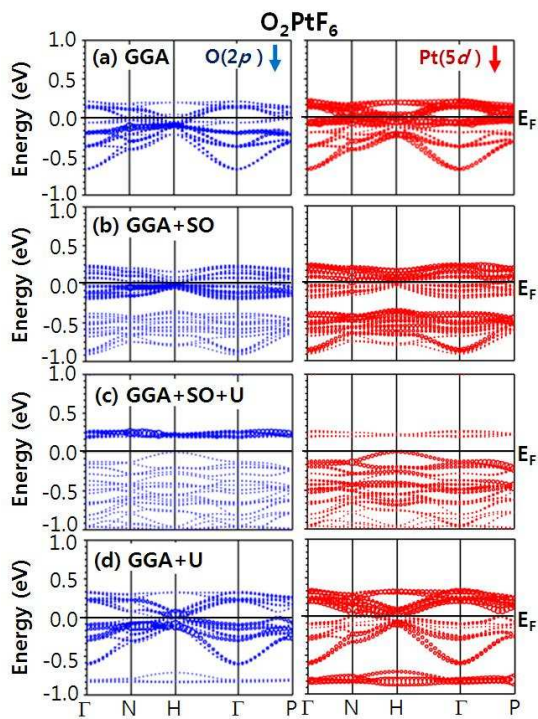}\\
\caption{(Color online)
(a) Spin $\downarrow$ band structures of O$_{2}$PtF$_{6}$ in the GGA.
The size of blue (red) dot denotes
the amount of O($2p$) (Pt(5$d$)) component in the wave function.
(b) The same as (a) in the GGA+SO.
(c) The same as (a) in the GGA+SO+$U$.
(d) The same as (a) in the GGA+$U$.
}
\label{fig6}
\end{figure}

Note, in Fig.~\ref{fig4}(a), that there exists substantial hybridization
effect between Pt(5$d$)-O($2p$) states,
even though Pt and O are not nearest neighbors (see Fig.\ref{fig1}).
In the case of Sr$_{2}$IrO$_{4}$, the SO interaction
not only splits degenerate $t_{2g}$ states but also reduces
the hybridization with neighboring O($2p$) states
because of its directionally non-polarized symmetric nature.\cite{bkim}
Similarly, in O$_{2}$PtF$_{6}$, Pt(5$d$)-O($2p$) hybridization
seems to be reduced a lot by considering the SO interaction.
Band structures in Fig.\ref{fig6} confirm this phenomenon.
The blue (red) dot denotes
the amount of O($2p$) (Pt(5$d$)) component in the wave function.
As shown in Fig.\ref{fig6}(a) and (d),
without inclusion of the SO interaction (GGA and GGA+$U$),
bands near $E_F$ are more dispersive.
In contrast, inclusion of the SO interaction in the GGA+SO and GGA+SO+$U$
reduces (i) the band dispersion,
(ii) the overlap between O($2p$)-Pt(5$d$) states,
and (iii) the  direction dependence of O($2p$) and Pt(5$d$) components,
as clearly seen in Fig.\ref{fig6}(b) and (c).
The SO split $\pi_{1/2}^{*}$ state in Fig.\ref{fig6}(b)
has a narrow band width of $W$=0.44 eV,
which is much smaller than that of the original antibonding $\pi$ states
($W$=0.84 eV) in Fig.\ref{fig6}(a).
Therefore, the SO interaction plays an important role in realizing
the Mott insulating state of O$_{2}$PtF$_{6}$, as in the case
of Sr$_{2}$IrO$_{4}$.

In conclusion, we have demonstrated that dioxygenyl magnet O$_{2}$PtF$_{6}$
is the first $2p$ electron Mott insulator induced by
combined effects of the Coulomb correlation and the SO interaction.
For O$_{2}$SbF$_{6}$, a typical 2$p$ molecular solid,
we have shown that the Coulomb correlation
of O($2p$) electrons drives the Mott insulating state.
The SO interaction in dioxygenyl O$_{2}^{+}$ in O$_{2}$SbF$_{6}$
is suppressed by the strong CF of neighboring F$^{-}$ ions,
as was corroborated by the quenched orbital moment of O$_{2}^{+}$.
For O$_{2}$PtF$_{6}$, the Mott insulating state is
obtained by considering both the Coulomb correlation
of O($2p$) and Pt(5$d$) electrons and the SO interaction of
of Pt(5$d$) states. The role of the SO interaction in O$_{2}$PtF$_{6}$
is similar to the case of Sr$_{2}$IrO$_{4}$,
which implies that O$_{2}$PtF$_{6}$ is also a typical
SO induced Mott system with J$_{eff}=1/2$.
The proposed aspects of strong SO interaction in O$_{2}$PtF$_{6}$
remain to be confirmed by further experiments:
(i) large orbital moment of Pt ion,
(ii) absence of the JT distortion of PtF$_{6}$ octahedron at low temperature.

This work was supported by the NRF (No.2009-0079947).
Helpful discussions with Beom Hyun Kim are greatly appreciated.


\begin{thebibliography}{99}
\bibitem{imada} M. Imada, A. Fujimori, and Y. Tokura,
      Rev. of Mod. Phys. {\bf 70}, 1039 (1998).
\bibitem{Kugel} K. I. Kugel and D. I. Khomskii, Sov. Phys. JETP
	 {\bf 37}, 725 (1973).
\bibitem{Bhkim10} B. H. Kim, H. Choi and B. I. Min,
	New J. Phys. {\bf 12}, 073023 (2010).
\bibitem{solovyev}{I. V. Solovyev,
	 New J. Phys.
     \textbf{10}, 013035 (2008).}
\bibitem{mkim} {M. Kim, B. H. Kim, H. C. Choi,
     and B. I. Min,
	 Phys. Rev. B
     \textbf{81}, 100409(R) (2010).}
\bibitem{kang} {J. -S. Kang, D. H. Kim,
     J. H. Hwang, J. Baik, H. J. Shin,
     M. Kim, Y. H. Jeong, and B. I. Min,
	 Phys. Rev. B
     \textbf{82}, 193102 (2010).}
\bibitem{bkim} {B. J. Kim, H. Jin, S. J. Moon,
     J. -Y. Kim, B. -G. Park, C. S. Leem,
     Jaejun Yu, T. W. Noh, C. Kim, S. -J. Oh,
     J. -H. Park, V. Durairaj, G. Cao, and
     E. Rotenberg,
	 Phys. Rev. Lett.
     \textbf{101}, 076402 (2008).}
\bibitem{moon} {S. J. Moon, H. Jin, K. W. Kim,
     W. S. Choi, Y. S. Lee, J. Yu, G. Cao,
     A. Sumi, H. Funakubo, C. Bernhard, and T. W. Noh,
	 Phys. Rev. Lett.
     \textbf{101}, 226402 (2008).}
\bibitem{pesin} {D. Pesin, and L. Balents,
	 Nature Phys.
     \textbf{6}, 376 (2010).}
\bibitem{watanabe} {H. Watanabe, T. Shirakawa,
     and S. Yunoki,
	 Phys. Rev. Lett.
     \textbf{105}, 216410 (2010).}
\bibitem{grill} {A. Grill, M. Schieber,
     and J. Shamir,
	 Phys. Rev. Lett.
     \textbf{25}, 747 (1970).}
\bibitem{disalvo} {F. J. Disalvo, W. E. Falconer,
     R. S. Hutton, A. Rodriguez, and J. V. Waszczak,
	 J. Chem. Phys.
     \textbf{62}, 2575 (1975).}
\bibitem{botkovitz} {P. Botkovitz, G. M. Lucier, R. P. Rao, and N. Bartlett,
	 Acta Chim. Slov.  \textbf{46}, 141 (1999).}
\bibitem{Attema05} {J. J. Attema, G. A. de Wijs, G. R. Blake, and
	R. A. de Groot, J. Am. Chem. Soc. \textbf{127}, 16325 (2005).}
\bibitem{kovacik} {R. Kov\'{a}\v{c}ik and C. Ederer, Phys. Rev. B
     \textbf{80}, 140411(R) (2009).}
\bibitem{winterlik} {J. Winterlik, G. H. Fecher, C. A. Jenkins, C. Felser,
     C. M\"{u}hle, K. Doll, M. Jansen, L. M. Sandratskii,
     and J. K\"{u}bler, Phys. Rev. Lett.
     \textbf{102}, 016401 (2009).}
\bibitem{Ylvi} {E. R. Ylvisaker, R. R. P. Singh, and W. E. Pickett,
	Phys. Rev. B \textbf{81}, 180405(R) (2010).}
\bibitem{nandy} {A. K. Nandy, P. Mahadevan,
     P. Sen, and D. D. Sarma, Phys. Rev. Lett.
     \textbf{105}, 056403 (2010).}
\bibitem{graudejus} {O. Graudejus and B. G. M\"{u}ller,
	 Z. anorg. allg. Chem.  \textbf{622}, 1076 (1996).}
\bibitem{FLAPW} M. Weinert, E. Wimmer, and A. J. Freeman,
	 Phys. Rev. B \textbf{26}, 4571(1982).
\bibitem{Blaha}{P. Blaha, K. Schwarz,
     G.K.H. Madsen, D. Kavasnicka,
     J. Luitz, WIEN2k  (Karlheinz Schwarz,
     Technische Universitat Wien, Austria, 2001).}
\bibitem{U}{V. I. Anisimov, I. V. Solovyev,
     M. A. Korotin, M. T. Czyzyk, and G. A. Sawatzky,
	Phys. Rev. B \textbf{48}, 16929 (1993).}
\bibitem{Uval}
	We have checked that electronic structures are consistent
	for $U$ values of $7 - 12$ eV.
	Employed $U$ value of 10 eV corresponds to the
	Coulomb interaction of 2$p$ electrons in O atoms
	or 5$d$ electrons in Pt atoms.
    Such large $U$ value would be screened by the intramolecular
    hybridization in O$_{2}^{+}$ or PtF$_{6}^{-}$,\cite{solovyev,takeshige}
	as displayed in the smaller energy separation
	between occupied and unoccupied states 
	of O($2p$) ($\Delta_H=2.55$ eV) or Pt(5$d$) ($\Delta_H=1.70$ eV)
	in Fig.\ref{fig4}(b). The screening
    by the intramolecular hybridization has also been confirmed
	in Ref.\cite{kang} for KO$_{2}$.
\bibitem{takeshige}{M. Takeshige, O. Sakai, and T. Kasuya,
	 J. Phys. Soc. Japn.
     \textbf{60}, 666 (1991).}
\bibitem{mconf}
	Due to the weak magnetic interactions, local electronic
    structures would not depend on the magnetic configuration
    under Mott insulating states.
\bibitem{kanzig} {W. K\"{a}nzig and M. Labhart,
     J. Phys. (Paris) 37, C7-39 (1976).}
\bibitem{shamir} {J. Shamir, and J. Binenboym,
	 Inorg. Chim. Acta
     \textbf{2}, 37 (1968).}
\bibitem{shimura} {T. Shimura, Y. Inaguma,
     T. Nakamura, M. Itoh, and Y. Morii,
	 Phys. Rev. B
     \textbf{52}, 9143 (1995).}
\bibitem{ibers}{J. A. Ibers, and W. C. Hamilton,
	 J. Chem. Phys.
     \textbf{44}, 1748 (1966).}

\end{thebibliography}
\end{document}